\documentclass[referee,a4paper,12pt,traditabstract]{swsc} 

\usepackage{graphicx}
\usepackage{txfonts}
\usepackage{subfigure}
\usepackage{epstopdf}
\usepackage{lineno}
\usepackage[authoryear,round]{natbib}
\usepackage[backref]{hyperref}
\usepackage{url}
\usepackage{tabularx}
\usepackage{rotating}
\hypersetup{colorlinks=true,citecolor=cyan,urlcolor=cyan,linkcolor=blue}

\begin{document}


   \title{Solar energetic particles and radio burst emission}

   
   \titlerunning{SEPs and solar radio bursts}

   \authorrunning{Miteva et al.}

   \author{R. Miteva
          \inst{1} 
          \and
          S. W. Samwel\inst{2}
           \and
          V. Krupar\inst{3,4,5}
          }

   \institute{Space Research and Technology Institute, Bulgarian Academy of Sciences (SRTI-BAS), 1113 Sofia, Bulgaria \\
              \email{\href{mailto:rmiteva@space.bas.bg}{rmiteva@space.bas.bg}}
         \and
             National Research Institute of Astronomy and Geophysics (NRIAG), 11421 Helwan, Cairo, Egypt \\
             \email{\href{mailto:samwelsw@nriag.sci.eg}{samwelsw@nriag.sci.eg}}
         \and
             Universities Space Research Association, 21046 Columbia, Maryland, USA
         \and
             NASA Goddard Space Flight Center, 20771 Greenbelt, Maryland, USA
         \and
             Institute of Atmospheric Physics CAS, 14131 Prague, Czech Republic\\
             \email{\href{mailto:vratislav.krupar@nasa.gov}{vratislav.krupar@nasa.gov}}
             }


 
  \abstract
   {We present a statistical study on the observed solar radio burst emission associated with the origin of in situ detected solar energetic particles. Several proton event catalogs in the period 1996$-$2016 are used. At the time of appearance of the particle origin (flare and coronal mass ejection) we identified radio burst signatures of types II, III and IV by inspecting dynamic radio spectral plots. The information from observatory reports is also accounted for during the analysis. The occurrence of solar radio burst signatures is evaluated within selected wavelength ranges during the solar cycle 23 and the ongoing 24. Finally, we present the burst occurrence trends with respect to the intensity of the proton events and the location of their solar origin.
   }        
   

   \maketitle

\section{Introduction}
     \label{S-Introduction} 

Solar energetic particles (SEPs) are protons, electrons and heavy ions energized during eruptive event at the Sun to keV$-$GeV energy \citep{{2006LRSP....3....2S},{2016LRSP...13....3D}}. If there is a possibility to escape, the particles propagate into the heliosphere along the interplanetary (IP) magnetic field lines. When these magnetic field lines pass over a particle detector aboard a satellite, it records the increasing particle intensity with time. The observed profile, however, depends on the driver productivity, the connectivity to the acceleration cite, the amount of seed particles from previous events, as well as eventual re-acceleration processes or particle loss in the IP medium. The most energetic SEPs ($>$500 MeV) can enter the terrestrial magnetosphere and the secondary products after interaction with atmospheric atoms can be detected by ground-based neutron monitors as ground level enhancements. In general, the SEP phenomena follows the overall solar cycle trend of eruptive phenomena \citep{2017SunGeo..12.0M}. From terrestrial point of view, however, the energetic particles carry some amount of latent risk. They are a source of the space radiation, which is dangerous to humans, and a reason for various spacecraft malfunction as well as accelerated material aging \citep{2007LRSP....4....1P}. Special efforts are carried out to improve the current capabilities to forecast/nowcast SEP events using different approaches and schemes.

The forecasting models are in general a multi-parameter solver based on various numerical techniques. The aim is to provide accurate particle arrival (and/or maximum intensity) prognosis with long warning time, few missed cases and low false alarm rate. The methods feed a mixture of input parameters into their models \citep{2009SpWea...7.4008L} or use: electrons as proton precursors \citep{SWE:SWE185}; location, intensity and steepness of the solar flare emission \citep{2011SpWea...9.7003N,2015SpWea..13..807N}; radio emission precursors \citep{2015ApJ...809..105W}; statistical probabilities \citep{2015SoPh..290..819T}, and evaluate a set of standard outputs. Validation procedures are imposed to cross-check the performance of each method.

Solar radio bursts were connected in past studies to in situ energetic particles, in terms of appearance of radio signatures \citep{1982ApJ...261..710K}, their duration \citep{2002JGRA..107.1315C} and spectral properties \citep{1990AN....311..379C}, with ongoing follow-up discussion in the literature. In the present analysis, we focus mainly on the presence or not of radio bursts in relationship to SEP events, their solar origin and strength.

The solar radio domain possesses unique (remote-sensing) diagnostic potential on drivers, environments, processes, particle distributions. The radio emission originates from different physical processes acting on populations of energetic electrons \citep{1985srph.book..177M}. Decimetric emission (dm, in radio wavelength notation) originates in the low corona and is due to gyrosynchrotron, gyroresonance, and/or bremsstrahlung, whereas in the higher corona (metric, m, to Decametric, Dm) and the IP space (Hectometric, Hm, to kilometric, km, range) the plasma mechanism dominates. Historically, solar radio emission features \citep{{1963ARA&A...1..291W},{2008SoPh..253....3N},{2008A&ARv..16....1P}} are separated into several classes or types, according to their appearance in the dynamic radio spectrum. Among other criteria used are the burst duration, drift rate, emission mechanism, accelerator, fine structures. In short, nowadays it is accepted that: type I bursts denote noise storms, type II $-$ shock waves \citep{1985srph.book..333N}, type III $-$ electron beams \citep{1985srph.book..289S}, type IV $-$ trapped electrons \citep{{1985srph.book..361S},{1986SoPh..104...19P}} and type V $-$ continuum following type III bursts \citep{1985srph.book..289S}. The space weather relevance of the different burst types is discussed in \cite{2004LNP...656...49W}.
     
Previous studies on the relationship between in situ particles and radio emission signatures can be divided into two main groups, depending on the approach. Namely, the analysis may initiate from a list of given radio signatures followed by the search of in situ particles; for studies in the previous solar cycle (SC) see, e.g., \cite{{2004ApJ...605..902C},{2008AnGeo..26.3033G}}. These earlier reports show larger association rate when type II bursts are present both in the m and Dekametric-Hectometric (DH) range, extending from about 50\% (for eastern origin) to 90\% (western origin), based on partial SC analysis, whereas about 60\% is reported for a larger sample, respectively. Recently, \cite{2015ApJ...809..105W} evaluated that DH type III associated with II bursts are an important asset for SEP predicting purposes, with success rate similar of the human forecasters. Furthermore, \cite{2017Ap&SS.362...56P} showed that 65\% of the m-to-DH type II are associated with SEP events.

Alternatively, a study may start with a particle catalog and then evaluate the appearance or not of specific radio bursts. Earlier works report large occurrence rates of metric ($\sim$90\%) and DH type II bursts (100\%) when associated with strong or major proton events, see e.g., \cite{{2003GeoRL..30.8013G},{2004ApJ...605..902C},{2007ApJ...658.1349C}}. Interestingly, the same rates are reported for metric (100\%) and DH type III bursts ($\sim$90\%) by \cite{{2002JGRA..107.1315C},{2003GeoRL..30.8018M}}, respectively, again in relation to protons with large intensity. In a comprehensive analysis over SC23, \cite{2013CEAB...37..541M} reported that the association between $\sim$25 MeV proton events and the DH type III is always present, followed by type m and DH-II (75\%) and m-IV (about 60\%) bursts, see also their Table 4. Compared to the longitude of the proton and radio emission-producing flare, an increased occurrence rate is observed for the type II (reaching 88\% for DH and 94\% for m-II) and m-IV bursts (75\%) when originating at eastern helio-longitudes compared to those at the west (71\% for m and DH IIs, and 53\% for IVs, respectively). Recently, \cite{2016JSWSC...6A..42P} presented the association between proton events and solar radio bursts of type II in the DH range to be 74\% (with 5\% uncertain), 83\% with type III and 35\% (with additional 7\% noted as probable) with type IV bursts, using event sample during 1997$-$2013.

The present work adds to the latter studies. The aim of our analysis is to identify all radio burst types II, III and IV related to flares and CMEs that are related to the parent activity of SEP events during the period 1996$-$2016. In addition, we investigate the solar radio burst sample with respect to the longitudinal dependence of the solar origin location as well as a function of the peak intensity of the protons, divided into strong and weak. Furthermore, different combination of solar radio burst types in terms of the identification of the SEP origin are examined which could be potentially explored in future forecasting schemes. Finally, the trends observed over SC23 and in the ongoing SC24 in view of previous reports are discussed.

\section{Data}
\label{S-Data}

\subsection{Proton events and their solar origin}
\label{S-pr_list}

For the purpose of our study, we prepare a proton event list independent on satellite, instrument or energy range. Namely, a comparative analysis on the proton events reported by several proton catalogs is performed (based on energies between $\sim$10 and $\sim$70 MeV), as summarized in Table~\ref{T-cats}.

\begin{table}[t!]
\caption[]{Table of selected proton catalogs used to compile the generalized proton event list used in this study (pfu - proton flux unit).}
\label{T-cats}
\small
\vspace{0.01cm}
\begin{tabularx}{\textwidth}{lllll}
\hline
Catalog & Satellite/ & Energy   & Yearly   & Median peak  \\
name    & Instrument & coverage & coverage & proton intensity \\
\hline
SEPEM      & GOES        & 7.23$-$10.45 MeV & 1997$-$2013 & 15.5 (cm$^2$ sr s MeV)$^{-1}$ \\
           & \multicolumn{4}{l}{\url{http://dev.sepem.oma.be/help/event_ref.html}; \cite{2015SpWea..13..406C} } \\
GOES-NOAA  & GOES        & $>$10 MeV        & 1996$-$2016 & 64 pfu \\
           & \multicolumn{4}{l}{\url{https://umbra.nascom.nasa.gov/SEP/}} \\
GOES-SSE   & GOES        & $>$10 MeV        & 1996$-$2013 & 47.5 pfu \\
           & \multicolumn{4}{l}{\cite{2015SoPh..290..841D}} \\
GOES-SEP   & GOES        & $>$10 MeV        & 1996$-$2013 & 24.2 pfu \\
           & \multicolumn{4}{l}{\cite{2016JSWSC...6A..42P}} \\
IMP-8      & IMP-8; SOHO & $>$25 MeV        & 1997$-$2006 & 0.008 (cm$^2$ sr s MeV)$^{-1}$ \\
           & \multicolumn{4}{l}{\cite{2010JGRA..11508101C}} \\
Wind/EPACT & Wind/EPACT  & 19$-$28 MeV      & 1996$-$2016 & 0.0361 (cm$^2$ sr s MeV)$^{-1}$ \\
           & \multicolumn{4}{l}{\url{http://newserver.stil.bas.bg/SEPcatalog/}; \cite{2017SunGeo..12.0M}} \\
SEPServer  & SOHO/ERNE   & 55$-$80 MeV      & 1996$-$2015 & 0.005 (cm$^2$ sr s MeV)$^{-1}$ \\
           & \multicolumn{4}{l}{\url{http://server.sepserver.eu/}; \cite{2013JSWSC...3A..12V}} \\
\hline
\end{tabularx}
\end{table}

We consider a reported particle event by different instruments to be the same phenomena, if the reported onset (and/or peak) times are within one day and the reported peak intensities are either large or small, compared to the median value for the sample. The detailed methodology for the compilation of the proton list is described in \cite{2017JASTP...M} where it was considered for the events in SC23. For the preset work the events in SC23 are re-analyzed and the list is extended to the ongoing SC24. Note that no SEP events are identified during 2009. The final proton list used in our analysis comprises 512 individual proton events from 1996 to 2016. The adopted criteria for the identification of flares and CMEs as the proton parent phenomena are as described in \cite{2017SunGeo..12.0M}. Namely, the strongest in soft X-ray (SXR) emission flare\footnote{\url{ftp://ftp.ngdc.noaa.gov/STP/space-weather/solar-data/solar-features/solar-flares/x-rays/goes/ and https://solarmonitor.org/}} and widest and fastest CME\footnote{\url{https://cdaw.gsfc.nasa.gov/CME_list/}} prior the particle onset at Earth are selected. Taking into account the particle profile and the timing of the strongest DH type III bursts as a proxy for the particle escape, the initial solar origin identification is revised if considered necessary. With the so-chosen procedure we identified 342 flares and 399 CMEs as the parent activity of the in situ particles. Back-sided eruptions, instrument gaps and uncertain association reduce the number of flares/CMEs compared to the number of proton events. The final event list consists of 431 proton events with identified either flare or/and CME and is summarised in the additional online material.

One of the objectives of the present study is to investigate the location trends of the solar eruptions responsible for the solar burst occurrence. Thus, the helio-longitude of the radio burst-producing flares and CMEs is required. The reported helio-longitude of the flare active region (AR) is used to separate the sample into eastern and western events. When no flare is identified  (flare was either behind the limb, uncertain identification or there were multiple candidates), the identified CME measurement positioning angle (MPA) is used which is measured in counterclockwise direction in order to qualify the phenomena into eastern (0$-$180 degrees) and western (181$-$360 degrees) longitude. 

\subsection{Radio dynamic spectra}
\label{S-dyn_spectra}

\begin{table}[t!]
\caption[]{Table of solar observatories providing radio spectral data used in this study ordered from high to low frequencies (pd: previous day).}
\label{T-radio_data}
\small
\vspace{0.01cm}
\begin{tabularx}{\textwidth}{lllll}
\hline
Radio        & Max frequency & Max UT   & Yearly    & Monitoring type \\
observatory  & range, MHz    & coverage & coverage  & \\
\hline
Ondrejov     & 800$-$5000 & 08$-$18          & 1991$-$present & campaigns  \\
             & \multicolumn{4}{l}{\url{http://www.asu.cas.cz/~radio/info.htm}}\\
e-Callisto/  & 10$-$5000   & 00$-$24          & 2002$-$present & database/patrol\\
Phoenix      & \multicolumn{4}{l}{\url{http://soleil.i4ds.ch/solarradio/callistoQuicklooks/}}\\  
             & \multicolumn{4}{l}{\url{http://soleil.i4ds.ch/solarradio/data/1998-2009_quickviews/}}\\            
HiRAS        & 25$-$2500  & 18$^{\rm pd}-$12 & 1996$-$2016    & patrol \\
             & \multicolumn{4}{l}{\url{http://sunbase.nict.go.jp/solar/denpa/index.html}}\\
Culgoora     & 18$-$1800  & 20$^{\rm pd}-$08 & 1992$-$present & patrol\\
             & \multicolumn{4}{l}{\url{http://www.sws.bom.gov.au/World_Data_Centre/1/9}}\\
Green Bank   & 5$-$1100   & 11$-$24          & 2004$-$2015    & patrol (data gaps)\\
             & \multicolumn{4}{l}{\url{http://www.astro.umd.edu/~white/gb/index.shtml}}\\
Radio Monitoring & 0.02$-$1000 & 00$-$24  & 1997$-$present & database \\
             & \multicolumn{4}{l}{\url{http://secchirh.obspm.fr/select.php}}\\
ARTEMIS      & 20$-$600   & 05$-$16          & 1998$-$2013    & patrol\\
             & \multicolumn{4}{l}{\url{http://artemis-iv.phys.uoa.gr/Artemis4_list.html}}\\
IZMIRAN      & 25$-$270   & 06$-$12          & 2000$-$present & patrol \\
             & \multicolumn{4}{l}{\url{http://www.izmiran.ru/stp/lars/s_archiv.htm}}\\
Learmonth    & 25$-$180   & 23$^{\rm pd}-$10 & 2000$-$present & patrol\\
             & \multicolumn{4}{l}{\url{http://www.sws.bom.gov.au/World_Data_Centre/1/9}}\\
Bruny Island & 6$-$62     & 18$^{\rm pd}-$10 & 2004$-$2015    & patrol (data gaps)\\
             & \multicolumn{4}{l}{\url{http://www.astro.umd.edu/~white/gb/index.shtml}}\\
Wind/WAVES   & 0.02$-$14  & 00$-$24          & 1994$-$present & patrol \\
STEREO/WAVES & 0.02$-$16  & 00$-$24          & 2007$-$present & patrol \\
             & \multicolumn{4}{l}{\url{https://solar-radio.gsfc.nasa.gov/wind/data_products.html}}\\
\hline
\end{tabularx}
\end{table}

We start our analysis with the list of flares and CMEs identified as the origin of proton events. We identified the radio bursts to the best of our knowledge over a period of about one hour following the flare onset. This reference period is used as a guidance in the search for radio emission signatures although there are exceptions in the case of long-duration events. When no flare could be identified, the start time is considered to be one hour prior the first appearance of CME over the occulting disk.

We use dynamic radio spectra, namely plots of the radio frequency vs. time, where the radio intensity is given as a color-code. We focus on the available on-line quick-look plots provided by various ground-based and space-borne radio observatories. Information on the solar-dedicated radio instruments used in our analysis, their coverage in frequency, UT, and years is shown in Table~\ref{T-radio_data}. In addition, the link to the repository of radio plots is provided. Only large periods of data gaps are explicitly noted, but all sources suffer from occasionally missed events. This is why, we inspected, where possible, all available plots for a given event from our list.
 
\subsection{Radio observatory reports}
\label{S-reports}

Occasionally, the quick-look radio spectral plots are of poor quality or the instrument sensitivity is low for specific frequency range. In addition, numerous terrestrial interferences compromise the radio burst continuation and intensity as depicted on the dynamic spectrum. Since the period of analysis extends after 2009, some of the observatories no longer exist (or use different instruments) and/or their data is no longer provided on-line. This is the reason why we compare the radio burst identification as performed by us with those provided by observers on duty at each radio observatory. As a verification of the burst identification, all radio observatory reports, where provided, are collected.

The list of observatories and databases reporting the occurrence of solar radio bursts is given in Table~\ref{T-radio_reports}. In addition to the NOAA-comprehensive database reports, the listings made individually by each radio observatory are checked if still provided at their web-sites. This is done since omissions were found in some database listings. Occasionally, some observatories tend to report types II and IV omitting the numerous type IIIs. Other observatories provide explicit burst identification only for specific days or only for outstanding burst appearances (selected events). We cross-checked the numerous observatory reports in order to perform as comprehensive coverage as possible.

\begin{table}[t!]
\caption[]{Temporal coverage and link to the radio observatory reports used in this study.}
\label{T-radio_reports}
\small
\vspace{0.01cm}
\begin{tabularx}{\textwidth}{ll}
\hline
Coverage  & Link/Reference  \\
\hline
\multicolumn{2}{c}{NOAA} \\
\multicolumn{2}{c}{\url{ftp.ngdc.noaa.gov/STP/space-weather/solar-data/solar-features/solar-radio}} \\
1960$-$2010 & \url{/radio-bursts/fixed-frequency-listings/} \\
1967$-$2011 & \url{/radio-bursts/reports/spectral-listings/} \\
2000$-$2011 & \url{/radio-bursts/tables/spectral-sgd/} \\    
1955$-$2009 & \url{ftp.ngdc.noaa.gov/STP/SOLAR_DATA/SGD_PDFversion/} \\
1996$-$present & \url{ftp.ngdc.noaa.gov/STP/swpc_products/daily_reports/solar_event_reports/} \\
\multicolumn{2}{c}{Phoenix/e-Callisto} \\
1998$-$2011 & \url{http://soleil.i4ds.ch/solarradio/} \\
\multicolumn{2}{c}{HiRAS} \\
1994$-$2011 & \url{http://sunbase.nict.go.jp/solar/denpa/spe_summary/} \\
\multicolumn{2}{c}{Culgoora/Learmonth} \\
1992$-$present & \url{http://www.sws.bom.gov.au/World_Data_Centre/1/9}  \\
\multicolumn{2}{c}{OSRA} \\
1990$-$2007 & \url{http://www.aip.de/osra/data/montab/} \\
\multicolumn{2}{c}{ARTEMIS (selected events)} \\
1998$-$2013 & \url{http://artemis-iv.phys.uoa.gr/DataBaseForWeb/data_set.htm}  \\
\multicolumn{2}{c}{IZMIRAN (selected events)} \\
1996$-$present & \url{http://www.izmiran.ru/stp/lars/MoreSp.html} \\
\multicolumn{2}{c}{Green bank/Bruny Island (selected events)} \\
2004$-$2015 & \url{http://www.astro.umd.edu/~white/gb/index.shtml} \\
\multicolumn{2}{c}{Wind/WAVES (selected events)} \\
1994$-$2015 & \url{https://solar-radio.gsfc.nasa.gov/wind/data_products.html} \\
\hline
\end{tabularx}
\end{table}

The same notation for the type II, III, IV bursts is adopted as introduced in \cite{2013CEAB...37..541M}. Namely, when a given burst is visually identified on the spectra we report the occurrence by their roman number starting by II, III and ending with IV. The notation is independent on the actual chronology of occurrence and is adopted here for clarity. We re-evaluated the events from SC23 using additionally found radio plots and observatory reports. There are occasional differences from the first identifications by \cite{2013CEAB...37..541M}, e.g., for the case of type II and IV bursts. This could be due to the fact that type II bursts are often difficult to recognize among the other emission. For the case of type IV burst, we focus on all broad-band and long-lasting radio emission occurring in the range from high to low frequencies. Occasionally, they are reported as decimetric continuum or continuum by observatory reports, whereas the notation `IV' is used by us.

Uncertain identification are denoted by `?' after the roman number of the burst type. When no images could be found by us for the specific event (or time period), but information for the occurrence of the radio bursts is given in observatory reports, the roman number is given in squared brackets. When no radio bursts could be identified by us on the plots (due to insufficient quality of the image, radio interference issues, low instrument sensitivity, etc.) or none are reported, we use `-' sign. If the identified by us radio burst was also confirmed by the observatory reports, it is noted solely by its roman number. Finally, we denote with `no data' in the table when neither spectral plots nor observatory reports could be found for a given proton event (in each radio wavelength range). In summary, the gaps are more frequent in the high frequency ranges, since the data coverage there is not complete.

In the table in the additional online material are given all visual identifications for the radio burst types in relation to the SEP solar origin as described above. The events are listed chronologically, where the time is the onset time of the related solar flare or, alternatively, the time of first appearance of the related CME, when no flare could be identified.

\section{Results}
\label{S-Results} 

We present the radio burst occurrence over five wavelength (frequency) ranges, namely: dm (3$-$1$\,$GHz); dm$-$m (1$-$0.3$\,$GHz); m (300$-$100$\,$MHz); m$-$Dm (100$-$30$\,$MHz); and Dm$-$Hm$-$km (30$-$0.01$\,$MHz), as also described in the table in the additional online material (for actual data coverage see Table~\ref{T-radio_data}). The radio burst occurrence is noted in the specific wavelength bin even though the burst may not cover the entire range. For comparative purpose, the number of radio bursts is given after being normalized to the specific event sample (e.g., Entire sample; Western or Eastern origin location sub-samples; strong or weak proton intensity sub-samples), and the ratios are reported in percentage. In addition, the uncertainty on each ratio is calculated according to error propagation rule. The latter results are very close (to within several percent) to the Poisson uncertainty (square root of the number of events in the sample). The data gaps over the entire event sample are most numerous in the highest frequency dm-range $-$ 83 cases, becoming fewer with decreasing frequency, namely to 67, 27, 10, and none in at Dm$-$Hm$-$km (noted as DH$-$km), respectively. Since the data gap is a fixed number in a given frequency range, the fraction on the plots is also constant for any of the burst type (plotted also in \%, as the ratio to the event number in the specific sample). The gaps are plotted in light-gray color on all histograms.

Since the proton sample is not observed by unique instrument, but composed from several data sets, it is not possible to perform statistical correlation analysis between the proton peak intensity and the solar origin characteristics. Instead, in order to provide additional information on the origin of the protons and solar radio bursts, we describe the distributions of flares and CMEs, related to a selected sub-sample in our study. 

\subsection{Overall trends}

\begin{figure}[t!]
\centerline{\includegraphics[width=\textwidth,clip=]{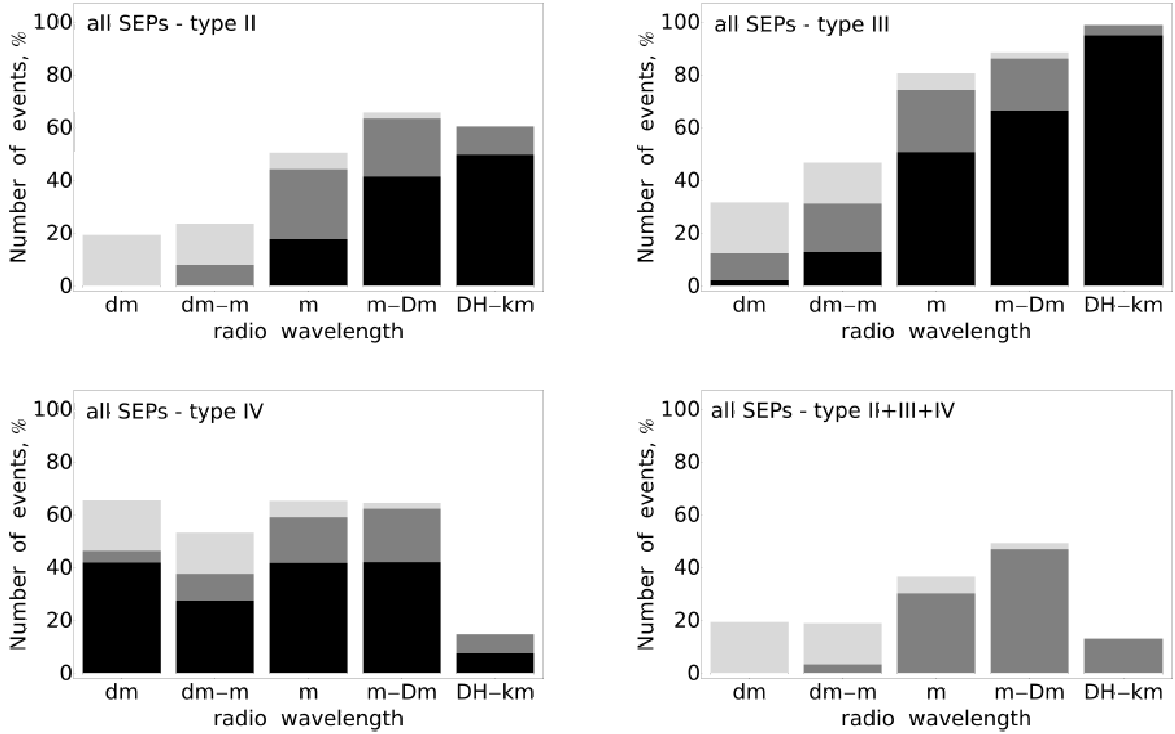}}      
\caption[]{\small Histograms of the percentage of radio burst types II, III, IV and a combination of all three types in association to SEP events (1996$-$2016) over several radio wavelength bins. Black color is for visually observed bursts on dynamic radio spectra, dark-gray for uncertain identification and observatory reports and light gray for data gaps. For the burst combination, the dark-gray color denotes random combination of visual, uncertain and observatory reports.}
    \label{F-All}
\end{figure}

The overall occurrence of radio burst types II, III, IV and their combination, noted as II+III+IV, in the period 1996$-$2016 is shown as histograms in Fig.~\ref{F-All}. There, the same plotting scheme is adopted as introduced by \citet{2013CEAB...37..541M}. Namely, the results are presented in five radio wavelength bins and each bin gives the normalized number of bursts to the entire number of events in the considered sample.

We start by counting the number of a given burst type in each of the wavelength ranges, denoting the level of confidence on the identification: visually confirmed, uncertain identification or only reported. When the overall sample is considered, the number of bursts in each wavelength bin is normalized to the entire sample size (namely 431 events for 1996$-$2016). Finally the ratios are plotted in percentage in Fig.~\ref{F-All} (as stacked histogram) whereas the values are explicitly summarized in Table~\ref{T-All}. 

\begin{table}[t!]
\centering
\caption[]{Table of the radio burst types occurrence in \% of the entire event sample (normalized to the sample size) with its uncertainty calculated as error propagation of the ratio. The results are given for the entire sample of events (independent on longitude or strength) over the entire period 1996$-$2016, and separately for SC23 (1996$-$2008) and SC24 (2009$-$2016). Each occurrence is the sum of the visual identifications/uncertain identifications/solely observatory reports, respectively, with their individual percentage contributions given explicitly in brackets. The percentage of the data gaps are not included here but their explicit event number is given in the text.} 
\label{T-All}
\footnotesize 
\begin{tabular}{llllll}
\hline
Radio       & \multicolumn{5}{c}{Radio wavelength} \\
burst       & dm          & dm$-$m       & m         & m$-$Dm    & DH$-$km \\
\hline
\multicolumn{6}{c}{Entire sample (1996$-$2016): 431 events}\\
II          & 0         & 8$^{\pm 1}$ (0/1/7) & 44$^{\pm 4}$ (18/1/25) & 63$^{\pm 5}$ (41/3/19) & 60$^{\pm 5}$ (49/4/7) \\
III         & 12$^{\pm 7}$ (2/1/9)   & 31$^{\pm 13}$ (13/2/16) & 74$^{\pm 26}$ (50/3/21) & 86$^{+14/-30}$ (66/2/18) & 99$^{+1/-34}$ (95/4/0) \\
IV          & 46$^{\pm 18}$ (42/1/3) & 37$^{\pm 15}$ (27/4/6)  & 59$^{\pm 22}$ (42/3/14) & 62$^{\pm 22}$ (42/3/17) & 14$^{\pm 8}$ (7/4/3) \\
II+III+IV   & 0         & 3$^{\pm 1}$  & 30$^{\pm 3}$  & 47$^{\pm 4}$  & 13$^{\pm 2}$   \\
\hline
\multicolumn{6}{c}{Entire sample in SC23 (1996$-$2008): 303 events}\\
II          & 0           & 10$^{\pm 2}$ (0/1/9) & 49$^{\pm 5}$ (21/0/28) & 65$^{\pm 6}$ (44/1/20) & 55$^{\pm 5}$ (45/1/9) \\
III         & 15$^{\pm 2}$ (2/1/12) & 32$^{\pm 4}$ (13/0/19) & 78$^{\pm 7}$ (52/1/25) & 89$^{\pm 7}$ (68/1/20) & 99$^{+1/-8}$ (96/3/0) \\
IV          & 51$^{\pm 5}$ (46/1/4) & 55$^{\pm 5}$ (45/2/8)  & 61$^{\pm 7}$ (40/4/17) & 64$^{\pm 6}$ (42/4/18) & 17$^{\pm 3}$ (12/4/1) \\
II+III+IV   & 0           & 4$^{\pm 1}$          & 33$^{\pm 4}$           & 48$^{\pm 5}$           & 13$^{\pm 2}$   \\
\hline
\multicolumn{6}{c}{Entire sample in SC24 (2009$-$2016): 128 events} \\
II          & 0                     & 3$^{\pm 1}$ (0/1/2)    & 32$^{\pm 6}$ (9/4/19)  & 59$^{\pm 9}$ (35/6/18)  & 73$^{\pm 10}$ (59/11/3) \\
III         & 7$^{\pm 2}$ (2/3/2)   & 29$^{\pm 5}$ (13/9/7)  & 65$^{\pm 9}$ (46/8/11) & 79$^{\pm 10}$ (61/5/13) & 98$^{+2/-12}$ (91/7/0) \\
IV          & 33$^{\pm 6}$ (30/2/1) & 52$^{\pm 8}$ (41/10/1) & 61$^{\pm 9}$ (45/11/5) & 58$^{\pm 8}$ (41/3/14)  & 17$^{\pm 4}$ (5/3/9) \\
II+III+IV   & 0                     & 2$^{\pm 1}$            & 23$^{\pm 5}$           & 44$^{\pm 7}$            & 13$^{\pm 4}$  \\
\hline
\end{tabular}
\end{table}

Different color notations are used on the plots for the different certainty of burst identifications. When a given radio burst was identified by us on the dynamic radio spectrum, we represent their fractional number in the specific wavelength range in black color. Uncertain identifications are shown in dark-gray color as well as the cases when the bursts were reported only by the listing issued by the radio observatories (for the latter cases no radio plots could be found). The light-gray code is used for the data gaps. The latter are fixed number in the corresponding frequency ranges for the different burst types. The sum of all colors represent the maximum occurrence rate for the given burst type. For the case requiring the simultaneous presence of all bursts, II+III+IV, the dark-gray denotes a random combination of visual identifications, uncertain cases and observatory reports.

Due to the continuity of the space-based observations, the DH-range contains the smallest fraction of gaps (light-gray sections). Moreover, the combination of two spacecraft, observing from several vantage points (L1 and the twin STEREO mission), improved the detectability of the radio emission features. On the other hand, in the m$-$DH range, we had to rely the most to observatory reports, compared to the other wavelength ranges. In addition to this, there is a poor global coverage (from ground) at frequencies reaching the ionospheric cutoff. We report burst types with largest amount of uncertainty (dark-gray sections) in the m and m$-$Dm range, due to overlapping with other bursts and/or being very faint and intermittent structures on the radio spectra).

In Table~\ref{T-All} are listed the exact percentages (sum of the black and dark-gray section) with their uncertainties. The individual contribution, again in percentages, are shown in brackets, separately for the visually identified events, for the visually identified but with some amount of uncertainty, and finally for those taken from observatory reports. The roundoff uncertainty is of the order of 1\%.

For the entire sample (1996$-$2016) the occurrences of type II and III increase with the increase of radio wavelength, the maximum being observed at m$-$Dm and DH$-$km range, respectively. Namely, type II bursts reach 63\%, whereas type IIIs are nearly always present, 99\%. In contrast, in DH$-$km wavelength range, type IVs drop to 14\%, whereas their maximum value is in the m$-$Dm range, reaching up to 62\%. In addition, type IIs are almost absent in the low corona (dm-range), where IVs are most frequent radio burst type. Finally, the occurrence rate of the simultaneous presence of all three burst types is presented, which upper limit cannot surpass the maximum occurrence rate of the individual solar burst in the given wavelength bin.

Although the results over the entire period are presented both graphically and in Table~\ref{T-All}, we list in addition the results for the SC23 and the ongoing SC24. The tendencies in either SC is reminiscent to those for the entire sample, within the uncertainties. However, for the detailed comparison between the SCs the same number of years need to be used.

\begin{figure}[!t]
\centerline{\includegraphics[width=\textwidth,clip=]{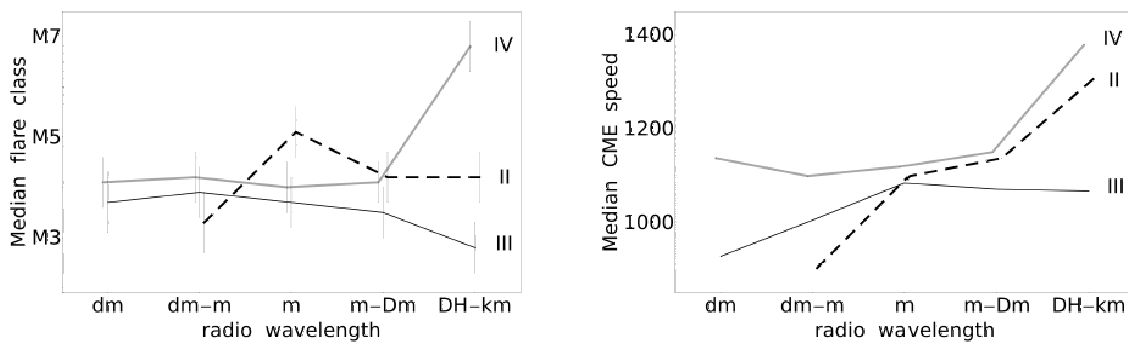}} 
\caption[]{Plot of the flare class (in W$\,$m$^{-2}$, on the left) and CME speed (in km$\,$s$^{-1}$, on the right) of samples associated with different radio burst types vs. the radio wavelength in the period 1996$-$2016. No restriction with respect to helio-longitude or proton intensity is imposed. Type II behavior is shown with black dashed line, type III $-$ with think black line, type IV $-$ with thick gray line.}
    \label{F-plot_all}
\end{figure}

Finally, the characteristics of the solar origin are inspected, namely we calculated the median values of the SXR flare class, CME projected speed and angular width (AW), related to radio bursts (and SEP events). We focus on the properties of the solar origin for each burst type separately. For the entire sample (neither limitation to the longitude of the solar origin, nor on the proton event intensity) the median value for the flare class is M2.8, for the CME projected speed $-$ 1055 km$\,$s$^{-1}$ and 360 degrees for the AW, respectively. The corresponding trends for the flares/CMEs when related to the different burst type samples (II, III and IV) over the different wavelength ranges are shown in Fig.~\ref{F-plot_all}. The error bars denote one standard deviation on the given sub-sample. Since the flare class covers more than 3 orders of magnitude, we calculate the uncertainty in log-scale. For the CMEs, the uncertainties are large, of the order of $\gtrsim$500 km$\,$s$^{-1}$ and are not shown on the plots for clarity purpose. The AW trend is not explicitly given, since the median values are very close to a full-halo CME in all wavelength bins.

There is very weak, if any, dependence of the flare class with respect to the wavelength of the accompanied three solar radio burst with the exception of the DH$-$km range, where the trends diverge (Fig.~\ref{F-plot_all}, left plot). Within the uncertainty the trends are relatively flat with marginal increase of the flare class related to m-type IIs. For the CME projected speed (Fig.~\ref{F-plot_all}, on the right), we obtain an increasing trend with radio wavelength for type II and IV bursts, wheres for type III some flattening is noticed at the longest wavelengths. In overall, larger flare class and CME speed (in median values) are accompanying type II and IV bursts compared to those related to type IIIs.

\subsection{Longitude trends}

\begin{figure}[!t]
\centerline{\includegraphics[width=\textwidth,clip=]{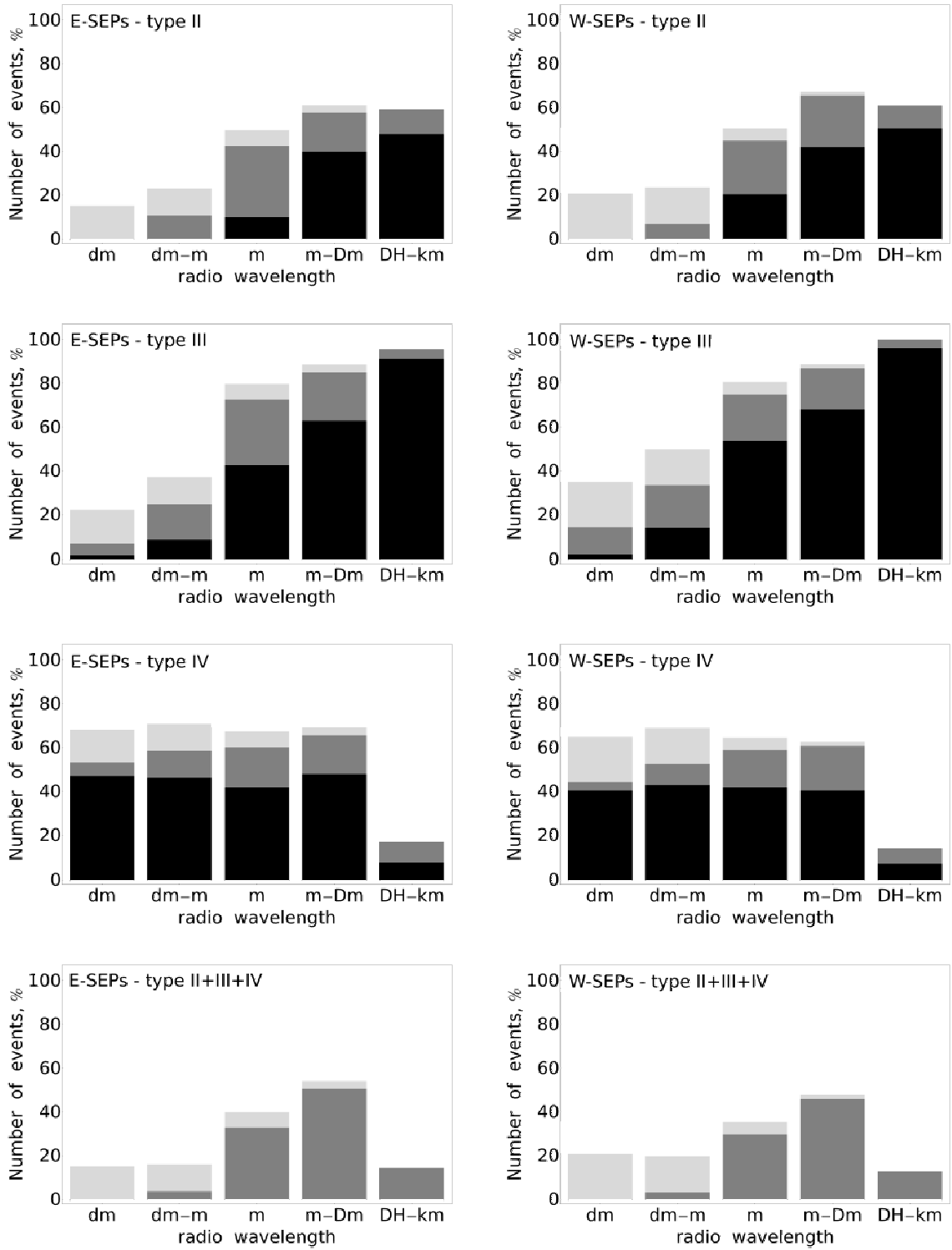}}               
\caption[]{Radio burst type occurrence related to proton events with Eastern (plots on the left, values normalized to 113) and Western (on the right, normalized to 315) origin. Color-code as in Fig.~\ref{F-All}.}
    \label{F-E-W}
\end{figure}

Here, the solar radio burst occurrence is evaluated separately for the Western and Eastern originating flares/CMEs. The number of Western events is 315/431 (73\%), whereas the fraction of eastern origin events is 113/431 (26\%). Three events are with uncertain origin location and thus are removed from the analysis. Once again, the radio burst occurrence is normalized over the total number of events in each category. The values for the different bursts are visually presented in Fig.~\ref{F-E-W} and also summarized in Table~\ref{T-long}. The data gaps, distributed with radio wavelength are 65/52/18/6/0 events with Western origin and 17/14/8/4/0 events from the East, respectively, shown only in the plots in \%.

In shape, all distributions follow the behavior of the entire event sample for the given burst type (compare with Fig.~\ref{F-All}). Similarly, the observatory reports contribute the most to the type II and III bursts at m and m$-$Dm regimes. Overall, there are no statistically significant differences in the Western compared to Eastern samples, over the entire period of the analysis (1996$-$2016). Some marginal differences are noted for Eastern m$-$Dm type IIs in SC24.

\begin{table}[t!]
\centering
\caption[]{Table of the radio burst types occurrence in \% related to Western and Eastern eruptive events in the three periods of interest. See Table~\ref{T-All} for further explanations.}
\label{T-long}
\footnotesize 
\begin{tabular}{llllll}
\hline
Radio       & \multicolumn{5}{c}{Radio wavelength} \\
burst       & dm          & dm$-$m       & m         & m$-$Dm    & DH$-$km \\
\hline
\multicolumn{6}{c}{Western origin sample (1996$-$2016): 315 events}\\
II          & 0           &  7$^{\pm 2}$ (0/1/6) & 44$^{\pm 5}$ (20/1/23) & 65$^{\pm 6}$ (42/3/21) & 61$^{\pm 6}$ (50/4/7) \\
III         & 14$^{\pm 2}$ (2/2/10) & 33$^{\pm 4}$ (14/3/16) & 75$^{\pm 6}$ (54/2/19) & 87$^{\pm 7}$ (68/2/17) & 100$^{-6}$ (96/4/0) \\
IV          & 44$^{\pm 4}$ (40/1/3) & 52$^{\pm 5}$ (43/4/5)  & 59$^{\pm 5}$ (42/4/13) & 60$^{\pm 6}$ (40/3/17) & 14$^{\pm }$ (7/3/4) \\
II+III+IV   & 0           & 3$^{\pm 1}$  & 30$^{\pm 3}$  & 46$^{\pm 5}$  & 13$^{\pm 2}$   \\
\multicolumn{6}{c}{Eastern origin sample (1996$-$2016): 113 events}\\
II          & 0           & 11$^{\pm 3}$ (0/0/11) & 43$^{\pm 7}$ (10/1/32) & 58$^{\pm 9}$ (40/2/16) & 59$^{\pm 9}$ (48/3/8) \\
III         & 7$^{\pm 3}$ (2/0/5)   & 25$^{\pm 5}$ (9/0/16) & 73$^{\pm 11}$ (42/4/27) & 85$^{\pm 12}$ (63/3/19) & 96$^{+4/-13}$ (91/5/0) \\
IV          & 53$^{\pm 8}$ (47/3/4) & 59$^{\pm 9}$ (46/4/9) & 60$^{\pm 9}$ (41/3/16)  & 65$^{\pm 10}$ (48/3/14) & 17$^{\pm 4}$ (8/6/3) \\
II+III+IV   & 0           & 4$^{\pm 2}$  & 33$^{\pm 6}$  & 50$^{\pm 8}$  & 14$^{\pm 4}$   \\
\hline
\multicolumn{6}{c}{Western origin sample in SC23 (1996$-$2008): 222 events}\\
II          & 0           & 9$^{\pm 2}$ (0/1/8) & 49$^{\pm 6}$ (24/0/25) & 65$^{\pm 7}$ (43/1/21) & 56$^{\pm 6}$ (47/1/8) \\
III         & 18$^{\pm 3}$ (3/1/14) & 34$^{\pm 5}$ (15/0/19) & 79$^{\pm 8}$ (56/1/22) & 90$^{\pm 9}$ (71/1/18) & 100$^{-9}$ (97/3/0) \\
IV          & 50$^{\pm 6}$ (46/1/3) & 53$^{\pm 6}$ (45/1/7)  & 57$^{\pm 4}$ (41/0/17) & 62$^{\pm 7}$ (40/3/19) & 13$^{\pm 3}$ (9/3/1) \\
II+III+IV   & 0           & 4$^{\pm 1}$  & 32$^{\pm 4}$  & 45$^{\pm 5}$  & 13$^{\pm 3}$   \\
\multicolumn{6}{c}{Eastern origin sample in SC23 (1996$-$2008): 78 events}\\
II          & 0           & 15$^{\pm 5}$ (0/0/15)  & 51$^{\pm 10}$ (13/0/38) & 66$^{\pm 12}$ (46/3/17) & 51$^{\pm 10}$ (40/1/10) \\
III         & 9$^{\pm 4}$ (1/0/8)    & 27$^{\pm 7}$ (6/0/21)   & 86$^{\pm 14}$ (43/8/35) & 91$^{+9/-15}$ (64/1/26) & 96$^{+4/-16}$ (94/2/0) \\
IV          & 58$^{\pm 11}$ (50/3/5) & 60$^{\pm 11}$ (44/3/13) & 60$^{\pm 11}$ (38/1/21) & 69$^{\pm 12}$ (51/4/14) & 16$^{\pm 5}$ (8/8/0) \\
II+III+IV   & 0           & 5$^{\pm 3}$           & 38$^{\pm 8}$           & 59$^{\pm 11}$  & 13$^{\pm 4}$   \\
\hline
\multicolumn{6}{c}{Western origin sample in SC24 (2009$-$2016): 93 events} \\
II          & 0                     & 3$^{\pm 3}$ (0/1/2)    & 34$^{\pm 7}$ (11/4/19)  & 66$^{\pm 11}$ (39/8/19) & 73$^{\pm 12}$ (58/12/3) \\
III         & 7$^{\pm 3}$ (1/4/2)   & 33$^{\pm 7}$ (13/12/8) & 66$^{\pm 11}$ (48/6/12) & 80$^{\pm 13}$ (61/4/15) & 100$^{-15}$ (94/6/0) \\
IV          & 30$^{\pm 6}$ (27/2/1) & 51$^{\pm 9}$ (38/12/1) & 62$^{\pm 10}$ (44/13/5) & 58$^{\pm 10}$ (41/3/14) & 15$^{\pm 4}$ (3/3/9)  \\
II+III+IV   & 0                     & 2$^{\pm 2}$            & 25$^{\pm 6}$            & 48$^{\pm 9}$            & 12$^{\pm 4}$  \\
\multicolumn{6}{c}{Eastern origin sample in SC24 (2009$-$2016): 35 events}\\
II          & 0                      & 0                      & 23$^{\pm 9}$ (3/3/17)   & 40$^{\pm 13}$ (26/0/14) & 75$^{\pm 19}$ (63/9/3)  \\
III         & 3$^{\pm 3}$ (3/0/0)    & 20$^{\pm 8}$ (14/0/6)  & 60$^{\pm 17}$ (40/11/9) & 72$^{\pm 19}$ (60/6/6)  & 94$^{+6/-23}$ (85/9/0) \\
IV          & 43$^{\pm 13}$ (40/3/0) & 56$^{\pm 15}$ (49/7/0) & 61$^{\pm 17}$ (49/6/6)  & 57$^{\pm 16}$ (40/3/14) & 21$^{\pm 8}$ (9/3/9) \\
II+III+IV   & 0                      & 0                      & 20$^{\pm 8}$            & 31$^{\pm 11}$           & 17$^{\pm 8}$ \\
\hline
\end{tabular}
\end{table}

\begin{figure}[!t]
\centerline{\includegraphics[width=\textwidth,clip=]{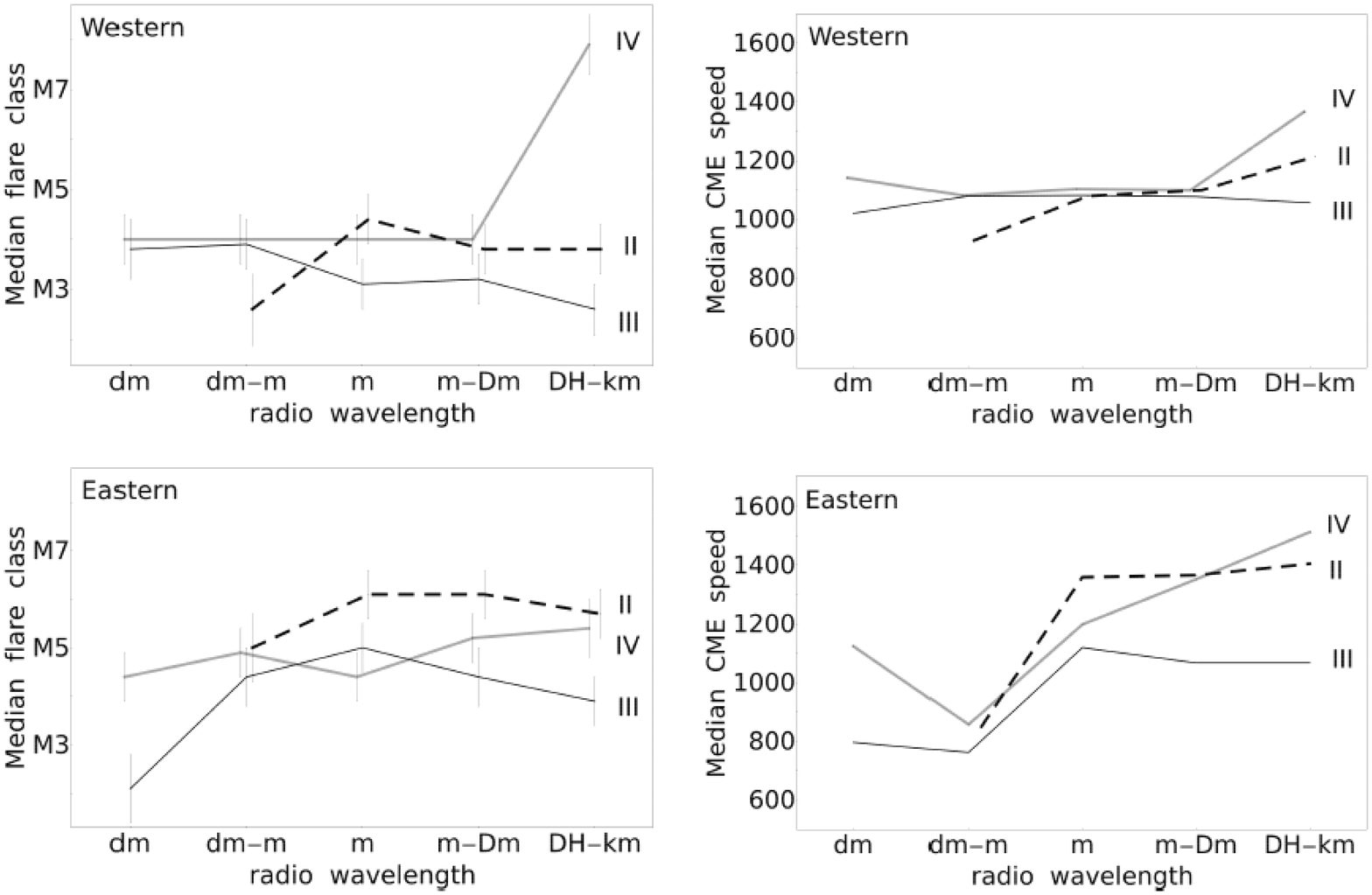}}   
\caption[]{Plot of the flare class (in W$\,$m$^{-2}$, on the left) and CME speed (in km$\,$s$^{-1}$, on the right) of samples associated with different solar radio burst types vs. the radio wavelength. Solar origin samples at western (eastern) helio-longitudes are given on the top (bottom). See Fig.~\ref{F-plot_all} for plotting style used.}
    \label{F-plot_W-E}
\end{figure}

Furthermore, in Fig.~\ref{F-plot_W-E} the median values of the flare and CMEs, related to the given event samples, are plotted over each wavelength range, given separately for Western (upper plots) and Eastern events (lower plots). The Western cases are again reminiscent to the results for the entire sample (Fig.~\ref{F-plot_all}), both for flare and CME trends, since the events with origin in the west consist of $\sim$70\% of all events. Thus, similar tendencies noticed in Fig.~\ref{F-plot_all}, are evident also for the Western events, Fig.~\ref{F-plot_W-E}, top plots.

For the dependence on the Eastern eruptive events with the radio wavelength (Fig.~\ref{F-plot_W-E}, bottom plots), we obtain that the flares responsible for II and III burst types have larger classes (in median values) compared to the Western values, with exception to type IVs in the DH$-$km range. The trend for the Eastern flares related to type II, III and IV bursts varies between M3.5$-$M6.5 with wavelength, although type dm-III bursts are accompanied with flares with median class of M2.5.

The behavior of the Eastern CMEs (median speed) as a function of longitude is flat for type IIs and IIIs with values at about 1400 and 1000 km$\,$s$^{-1}$, respectively, and has rising trend for type IVs in the m-to-km wavelengths. At lower wavelengths the trend shows a local minimum in the dm$-$m range.

\subsection{Proton intensity trends}

\begin{figure}[!t]
\centerline{\includegraphics[width=0.95\textwidth,clip=]{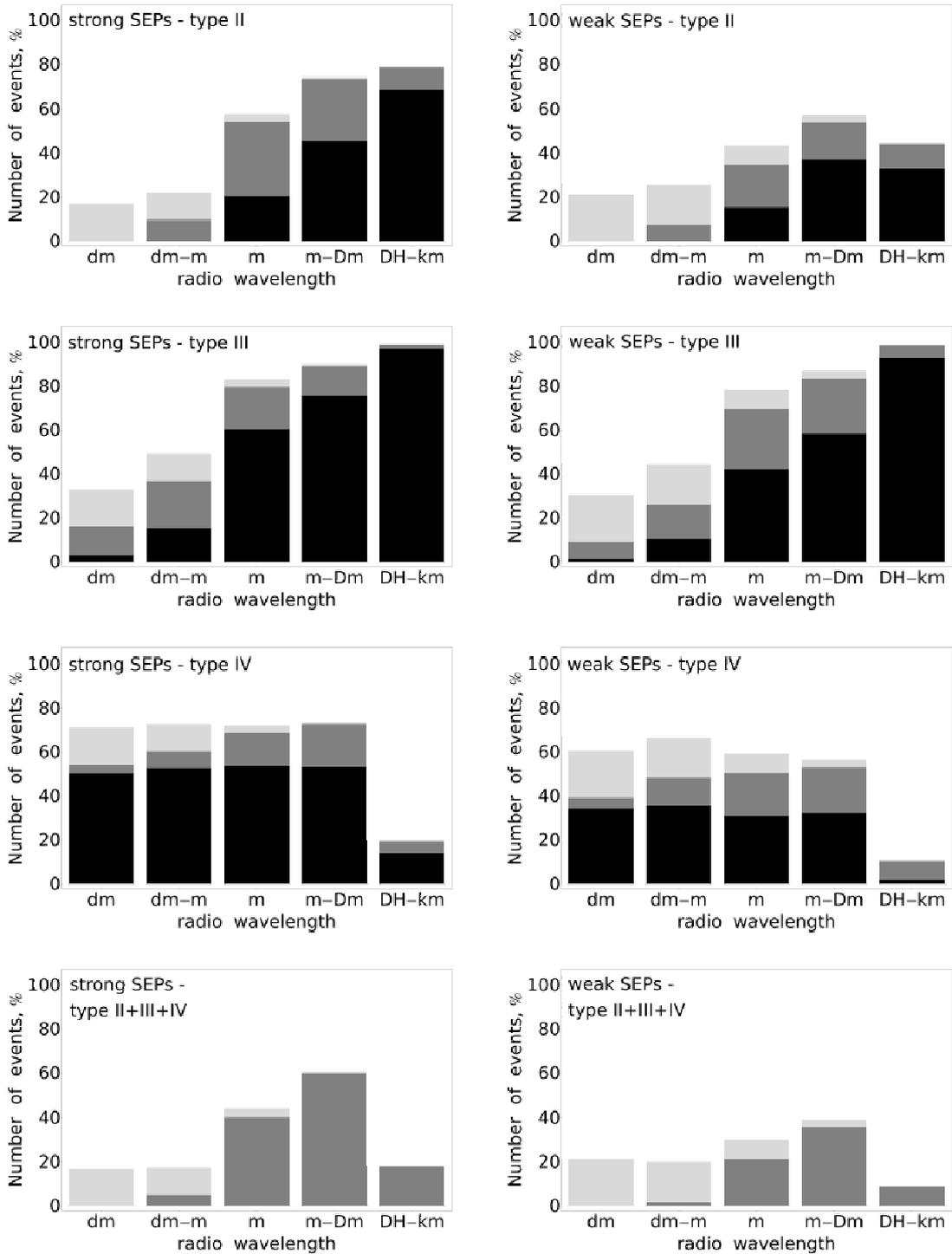}}              
\caption[]{Radio burst type occurrence related to strong (plots on the left, values normalized to 201) proton events and weak ones (on the right, normalized to 230). Color-code as in Fig.~\ref{F-All}.}
    \label{F-st-w}
\end{figure}

\begin{table}[t!]
\centering
\caption[]{Table of the radio burst types occurrence in \% related to strong and weak proton events in the three periods of interest. See Table~\ref{T-All} for further explanations.}
\label{T-int}
\footnotesize 
\begin{tabular}{llllll}
\hline
Radio       & \multicolumn{5}{c}{Radio wavelength} \\
burst       & dm          & dm$-$m       & m         & m$-$Dm    & DH$-$km \\
\hline
\multicolumn{6}{c}{Strong proton sample (1996$-$2016): 201 events}\\
II          & 0           & 9$^{\pm 2}$ (0/1/8) & 54$^{\pm 6}$ (20/2/32) & 74$^{\pm 8}$ (46/4/24) & 79$^{\pm 8}$ (69/4/6) \\
III         & 16$^{\pm 3}$ (3/2/11) & 37$^{\pm 5}$ (15/21/12) & 79$^{\pm 8}$ (60/4/15) & 89$^{\pm 9}$ (76/2/11) & 99$^{+1/-10}$ (97/2/0) \\
IV          & 54$^{\pm 6}$ (50/1/3) & 60$^{\pm 7}$ (53/3/4)   & 69$^{\pm 8}$ (54/3/12) & 72$^{\pm 9}$ (53/4/15) & 19$^{\pm 3}$ (14/4/1) \\
II+III+IV   & 0           & 10$^{\pm 2}$        & 40$^{\pm 5}$           & 60$^{\pm 7}$           & 18$^{\pm 3}$   \\
\multicolumn{6}{c}{Weak proton sample (1996$-$2016): 230 events}\\
II          & 0           & 7$^{\pm 2}$ (0/0/7) & 35$^{\pm 5}$ (15/1/19) & 54$^{\pm 6}$ (37/2/15) & 44$^{\pm 5}$ (33/4/7) \\
III         & 9$^{\pm 2}$ (1/1/7)   & 26$^{\pm 4}$ (10/2/14) & 70$^{\pm 7}$ (42/2/26) & 83$^{\pm 8}$ (58/2/23) & 99$^{+1/-9}$ (93/6/0) \\
IV          & 39$^{\pm 5}$ (34/3/2) & 48$^{\pm 6}$ (36/5/7)  & 50$^{\pm 6}$ (31/4/16) & 53$^{\pm 6}$ (32/3/18) & 11$^{\pm 2}$ (2/4/5) \\
II+III+IV   & 0           & 2$^{\pm 1}$        & 21$^{\pm 3}$           & 36$^{\pm 5}$           & 9$^{\pm 2}$   \\
\hline
\multicolumn{6}{c}{Strong proton sample in SC23 (1996$-$2008): 130 events}\\
II          & 0           & 12$^{\pm 3}$ (0/1/11) & 64$^{\pm 9}$ (26/0/38) & 76$^{\pm 10}$ (49/2/25) & 76$^{\pm 10}$ (66/1/9) \\
III         & 20$^{\pm 4}$ (4/0/16) & 36$^{\pm 6}$ (15/0/21) & 83$^{\pm 11}$ (63/2/18) & 92$^{+8/-12}$ (80/1/11) & 99$^{+1/-12}$ (98/1/0) \\
IV          & 60$^{\pm 9}$ (55/0/5) & 62$^{\pm 9}$ (54/2/6)  & 69$^{\pm 9}$ (52/0/17)  & 76$^{\pm 10}$ (54/4/18) & 22$^{\pm 5}$ (17/5/0) \\
II+III+IV   & 0           & 6$^{\pm 2}$          & 45$^{\pm 7}$           & 62$^{\pm 9}$            & 21$^{\pm 4}$   \\
\multicolumn{6}{c}{Weak proton sample in SC23 (1996$-$2008): 173 events}\\
II          & 0           & 9$^{\pm 2}$ (0/0/9) & 39$^{\pm 6}$ (18/0/21) & 58$^{\pm 7}$ (40/1/17) & 39$^{\pm 6}$ (30/1/8) \\
III         & 11$^{\pm 3}$ (1/1/9)  & 29$^{\pm 5}$ (14/19/16) & 75$^{\pm 9}$ (45/0/30) & 61$^{\pm 8}$ (60/1/0)  & 99$^{+1/-11}$ (95/4/0) \\
IV          & 45$^{\pm 6}$ (40/2/3) & 50$^{\pm 7}$ (38/2/10)  & 50$^{\pm 7}$ (31/1/18) & 55$^{\pm 7}$ (34/3/18) & 8$^{\pm 2}$ (2/4/2) \\
II+III+IV   & 0           & 3$^{\pm 1}$        & 24$^{\pm 4}$           & 38$^{\pm 5}$           & 7$^{\pm 9}$   \\
\hline
\multicolumn{6}{c}{Strong proton sample in SC24 (2009$-$2016): 71 events}\\
II          & 0                     & 4$^{\pm 2}$ (0/1/3)     & 38$^{\pm 9}$ (10/4/24)  & 72$^{\pm 13}$ (41/7/24) & 82$^{\pm 15}$ (73/8/1) \\
III         & 7$^{\pm 4}$ (1/4/3)   & 38$^{\pm 9}$ (17/10/11) & 74$^{\pm 13}$ (55/8/10) & 83$^{\pm 15}$ (68/4/11) & 99$^{+1/-17}$ (96/3/0) \\
IV          & 43$^{\pm 9}$ (41/1/1) & 58$^{\pm 11}$ (51/6/1)  & 69$^{\pm 13}$ (58/8/3)  & 67$^{\pm 13}$ (52/4/11) & 13$^{\pm 5}$ (8/1/4) \\
II+III+IV   & 0                     & 3$^{\pm 2}$             & 31$^{\pm 8}$            & 55$^{\pm 11}$           & 13$^{\pm 4}$  \\
\multicolumn{6}{c}{Weak proton sample in SC24 (2009$-$2016): 57 events}\\
II          & 0                     &  0                      & 23$^{\pm 7}$ (7/4/12)   & 43$^{\pm 10}$ (28/4/11) & 61$^{\pm 13}$ (42/14/5)  \\
III         & 4$^{\pm 3}$ (2/2/0)   & 18$^{\pm 6}$ (9/7/2)    & 54$^{\pm 12}$ (35/7/12) & 72$^{\pm 15}$ (53/5/14) & 98$^{+2/-18}$ (86/12/0) \\
IV          & 22$^{\pm 7}$ (18/4/0) & 44$^{\pm 11}$ (28/16/0) & 52$^{\pm 12}$ (29/14/9) & 46$^{\pm 11}$ (26/2/18) & 19$^{\pm 6}$ (0/5/14) \\
II+III+IV   & 0                     & 0                       & 14$^{\pm 5}$            & 30$^{\pm 8}$            & 14$^{\pm 5}$ \\
\hline
\end{tabular}
\end{table}

Here it is investigated the difference of the radio burst occurrence when related to strong and weak proton intensity events. Firstly, we separate the proton sample according to the median value of the peak proton intensity. The sample of strong proton events (namely, with values larger than the median) contains 201/431 (47\%) of the entire sample, whereas the weak events are 230/431 (53\%). These numbers are used for each normalization procedure. In addition, the data gaps in each respective radio wavelength bin are 34/25/7/2/0 for strong and 49/42/20/8/0 for weak sample, respectively, shown only in the plots in \%.

In general, a larger fraction of the radio bursts (of any type) are observed in relation to strong proton events, compared to the weak proton sample, see Fig.~\ref{F-st-w} (and also in Table~\ref{T-int}). This is most clearly noticeable for type II, followed by IV and III, with the exception of DH$-$km range where the occurrence of type IIIs is the same for strong and weak proton events. The exact values in SC23 and the ongoing SC24 are again listed in Table~\ref{T-int}.

The results reflect the different energetics of the events, namely larger energy release for the strong event sample compared to weak ones. In terms of their solar origin characteristics, we obtain that the median flare class for the strong proton sample is M5.4 and median CME speed is 1335 km$\,$s$^{-1}$. In contrast, for the weak proton sample the result is M1.4 and 850 km$\,$s$^{-1}$ for the respective solar origin. 

\begin{figure}[!t]
\centerline{\includegraphics[width=\textwidth,clip=]{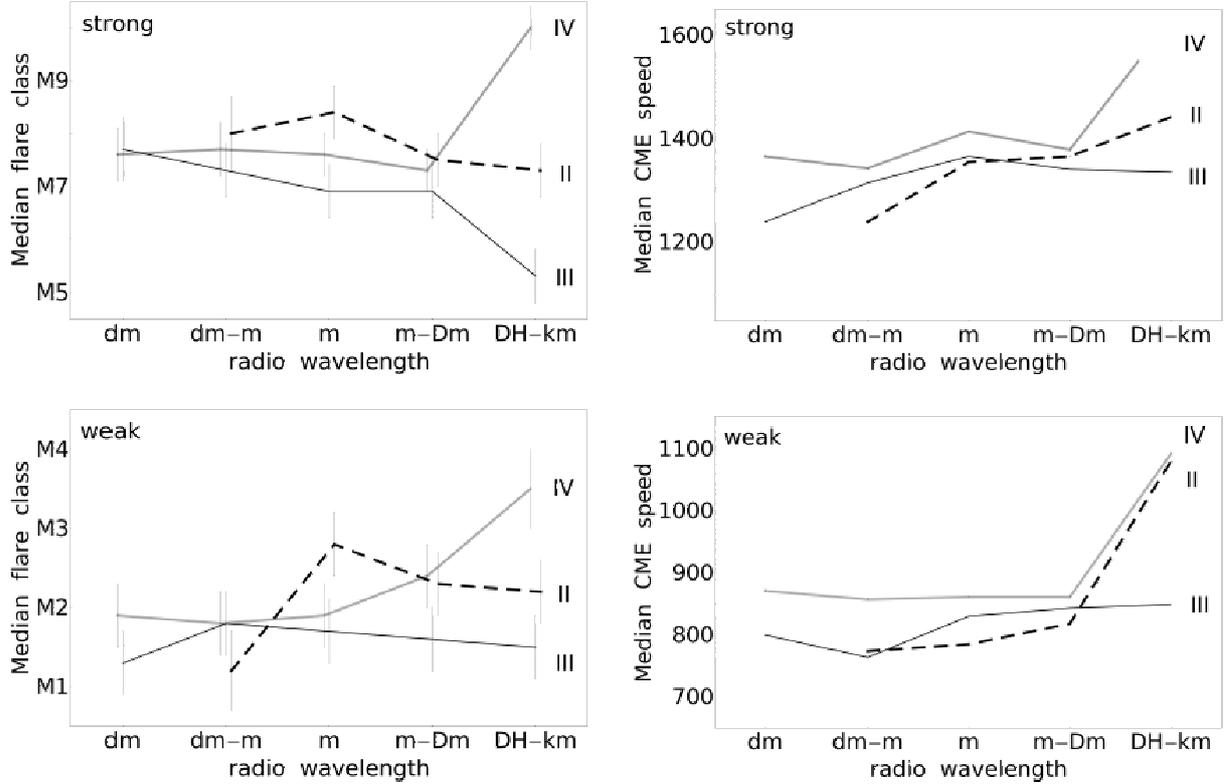}}    
\caption[]{Plot of the flare class (in W$\,$m$^{-2}$, on the left) and CME speed (in km$\,$s$^{-1}$, on the right) of proton samples associated with different radio burst types vs. the radio wavelength (1996$-$2016). Proton samples are separated in intensity as strong (weak), given on the top (bottom). See Fig.~\ref{F-plot_all}for plotting style used.}
    \label{F-plot_st-w}
\end{figure}

This can be inspected on the plots in Fig.~\ref{F-plot_st-w} where the flare/CME trends are explicitly given for strong (upper plots) and weak (lower plots) cases. The median flare class values, responsible for strong proton intensities, in relation to type II, III and IV show a trend similar to the entire and/or Western sample, diverging at longest wavelengths from $\sim$ M5 for type IIIs to $\sim$X1 for type IVs. In contrast, for the weak sample, the values drop at lower flare classes for types II and III but reach maximum to M3.5 for DH$-$km type IVs. For the median CME speeds, it is noticed that for the strong events the rising trend is observed at much higher value, up to $\sim$1500 km$\,$s$^{-1}$, whereas weak events related to type IIs and IVs reach maximum of about 1100 km$\,$s$^{-1}$, in contrast to type IIIs that are accompanied with CMEs at $\sim$800 km$\,$s$^{-1}$.

\subsection{Combination of radio bursts}

\begin{table}[t!]
\centering
\caption[]{Table of the occurrence (in \%) of different combinations of radio burst types.}
\label{T-Comb}
\footnotesize 
\begin{tabular}{llllll}
\hline
Radio burst & \multicolumn{5}{c}{Radio wavelength} \\
type        & dm          & dm$-$m       & m         & m$-$Dm    & DH$-$km \\
\hline
\multicolumn{6}{c}{Entire sample (1996$-$2016): 431 events} \\
II+III      & 0            & 20$^{\pm 2}$ & 47$^{\pm 4}$ & 62$^{\pm 5}$ & 61$^{\pm 5}$  \\
II+IV       & 0            & 22$^{\pm 3}$ & 39$^{\pm 4}$ & 51$^{\pm 2}$ & 13$^{\pm 2}$  \\
III+IV      & 29$^{\pm 3}$ & 41$^{\pm 4}$ & 58$^{\pm 5}$ & 61$^{\pm 5}$ & 15$^{\pm 2}$ \\
\multicolumn{6}{c}{Entire sample in SC23 (1996$-$2008): 303 events} \\
II+III      & 0            & 27$^{\pm 3}$ & 54$^{\pm 5}$ & 66$^{\pm 6}$ & 55$^{\pm 5}$ \\
II+IV       & 0            & 31$^{\pm 4}$ & 45$^{\pm 5}$ & 53$^{\pm 5}$ & 13$^{\pm 2}$  \\
III+IV      & 36$^{\pm 4}$ & 49$^{\pm 5}$ & 61$^{\pm 6}$ & 65$^{\pm 6}$ & 14$^{\pm 2}$  \\
\multicolumn{6}{c}{Entire sample in SC24 (2009$-$2016): 128 events} \\
II+III      & 0           &  2$^{\pm 1}$ & 30$^{\pm 5}$ & 55$^{\pm 8}$ & 73$^{\pm 10}$ \\
II+IV       & 0           &  2$^{\pm 1}$ & 25$^{\pm 5}$ & 45$^{\pm 7}$ & 13$^{\pm 3}$  \\
III+IV      & 4$^{\pm 2}$ & 23$^{\pm 5}$ & 48$^{\pm 7}$ & 52$^{\pm 8}$ & 16$^{\pm 4}$  \\
\hline
\multicolumn{6}{c}{Western/Eastern origin sample (1996$-$2016): 315/113 events} \\
II+III      & 0/0         & 21$^{\pm 3}$/16$^{\pm 4}$ & 47$^{\pm 5}$/46$^{\pm 8}$ & 64$^{\pm 6}$/58$^{\pm 9}$ & 61$^{\pm 6}$/60$^{\pm 9}$    \\
II+IV       & 0/0         & 22$^{\pm 3}$/23$^{\pm 5}$ & 38$^{\pm 4}$/42$^{\pm 7}$ & 49$^{\pm 5}$/55$^{\pm 9}$ & 13$^{\pm 2}$/14$^{\pm 4}$  \\
III+IV      & 31$^{\pm 4}$/21$^{\pm 5}$ & 43$^{\pm 4}$/34$^{\pm 6}$ & 57$^{\pm 2}$/60$^{\pm 9}$ & 59$^{\pm 5}$/65$^{\pm 10}$ & 14$^{\pm 2}$/17$^{\pm 4}$  \\
\multicolumn{6}{c}{Western/Eastern origin sample in SC23 (1996$-$2008): 222/78 events}  \\
II+III      & 0/0         & 28$^{\pm 4}$/23$^{\pm 6}$ & 53$^{\pm 6}$/58$^{\pm 11}$ & 65$^{\pm 7}$/69$^{\pm 12}$ & 56$^{\pm 6}$/54$^{\pm 10}$  \\
II+IV       & 0/0         & 30$^{\pm 4}$/33$^{\pm 8}$ & 43$^{\pm 5}$/52$^{\pm 10}$ & 49$^{\pm 6}$/64$^{\pm 12}$ & 13$^{\pm 3}$/13$^{\pm 4}$   \\
III+IV      & 39$^{\pm 5}$/26$^{\pm 6}$ & 51$^{\pm 6}$/42$^{\pm 3}$ & 59$^{\pm 7}$/67$^{\pm 12}$ & 61$^{\pm 7}$/74$^{\pm 13}$ & 14$^{\pm 3}$/15$^{\pm 5}$   \\
\multicolumn{6}{c}{Western/Eastern origin sample in SC24 (2009$-$2016): 93/35 events}   \\
II+III      & 0/0         & 3$^{\pm 2}$/0 & 33$^{\pm 7}$/20$^{\pm 8}$ & 62$^{\pm 10}$/34$^{\pm 11}$ & 73$^{\pm 12}$/74$^{\pm 19}$  \\
II+IV       & 0/0         & 2$^{\pm 2}$/0 & 26$^{\pm 6}$/23$^{\pm 9}$ & 48$^{\pm 9}$/34$^{\pm 11}$ & 12$^{\pm 4}$/17$^{\pm 8}$ \\
III+IV      & 4$^{\pm 2}$/3$^{\pm 3}$ & 26$^{\pm 6}$/14$^{\pm 7}$ & 49$^{\pm 9}$/46$^{\pm 14}$ & 55$^{\pm 10}$/43$^{\pm 13}$ & 15$^{\pm 4}$/20$^{\pm 8}$ \\
\hline
\multicolumn{6}{c}{Strong/Weak proton sample (1996$-$2016): 201/230 events}    \\
II+III      & 0/0         & 19$^{\pm 3}$/20$^{\pm 3}$ & 55$^{\pm 6}$/40$^{\pm 5}$ & 73$^{\pm 8}$/53$^{\pm 6}$ & 79$^{\pm 8}$/45$^{\pm 5}$  \\
II+IV       & 0/0         & 20$^{\pm 3}$/24$^{\pm 4}$ & 47$^{\pm 6}$/32$^{\pm 4}$ & 62$^{\pm 7}$/41$^{\pm 5}$ & 18$^{\pm 3}$/9$^{\pm 2}$  \\
III+IV      & 31$^{\pm 4}$/27$^{\pm 4}$ & 43$^{\pm 6}$/39$^{\pm 5}$ & 65$^{\pm 7}$/51$^{\pm 6}$ & 70$^{\pm 8}$/53$^{\pm 6}$ & 19$^{\pm 3}$/10$^{\pm 2}$  \\
\multicolumn{6}{c}{Strong/Weak proton sample in SC23 (1996$-$2008): 130/173 events}    \\
II+III      & 0/0         & 27$^{\pm 5}$/27$^{\pm 4}$ & 64$^{\pm 9}$/47$^{\pm 6}$ & 75$^{\pm 10}$/59$^{\pm 7}$ & 76$^{\pm 10}$/39$^{\pm 6}$  \\
II+IV       & 0/0         & 29$^{\pm 5}$/32$^{\pm 5}$ & 55$^{\pm 8}$/37$^{\pm 5}$ & 65$^{\pm 9}$/44$^{\pm 6}$  & 21$^{\pm 4}$/6$^{\pm 2}$  \\
III+IV      & 40$^{\pm 7}$/32$^{\pm 5}$ & 49$^{\pm 8}$/48$^{\pm 6}$ & 68$^{\pm 9}$/56$^{\pm 7}$ & 75$^{\pm 10}$/57$^{\pm 7}$ & 22$^{\pm 5}$/8$^{\pm 2}$  \\
\multicolumn{6}{c}{Strong/Weak proton sample in SC24 (2009$-$2016): 71/57 events}      \\
II+III      & 0/0         & 4$^{\pm 2}$/0 & 38$^{\pm 9}$/19$^{\pm 6}$ & 69$^{\pm 13}$/37$^{\pm 9}$ & 83$^{\pm 15}$/61$^{\pm 13}$ \\
II+IV       & 0/0         & 3$^{\pm 2}$/0 & 31$^{\pm 8}$/18$^{\pm 6}$ & 55$^{\pm 11}$/32$^{\pm 9}$ & 13$^{\pm 4}$/14$^{\pm 5}$ \\
III+IV      & 6$^{\pm 3}$/2$^{\pm 2}$ & 31$^{\pm 8}$/12$^{\pm 5}$ & 58$^{\pm 11}$/37$^{\pm 9}$ & 62$^{\pm 12}$/39$^{\pm 10}$ & 14$^{\pm 5}$/19$^{\pm 6}$ \\
\hline
\end{tabular}
\end{table}

For completeness, we calculated the percentages for all combinations of radio bursts: II+III, II+IV, and III+IV. In this case, no separation is made on the method of identification (visual, uncertain or only reported), thus the result represents an upper limit for the occurrence. The requirement is the simultaneous presence of the given pair of burst types irrelevant to the occurrence of the other solar burst type. The results are explicitly given in Table~\ref{T-Comb} together with the uncertainties, calculated in a similar way as for the values in Table~\ref{T-All}.

As a rule, the occurrence rate for a given burst combination cannot surpass the maximum value obtained of any of the bursts in the combination. Thus, for all combinations including type II, the value is below $\sim$70$-$80\% in the DH$-$km range, whereas for combinations with type IV in the same wavelength range, the value is always low, since type IV are not frequent there. The largest value is obtained for DH$-$km II+III related to the strong sample, reaching 83\%. 

In overall, types II+III show rising trend with radio frequency, whereas types II+IV and III+IV show the slowly rise with a flattening, ending with an abrupt decrease at DH-wavelengths, that is due to the type IV spectral behavior. The statistically significant differences are obtained for the Strong/Weak event samples, more often in the dm$-$m, m and m$-$Dm ranges.

\subsection{On the origin of SEP events}
\label{S-Discussion}

Although flare acceleration is plausible over the entire range of interest (dm$-$km), the influence of CMEs to low-coronal emission (where type II observations are usually lacking) is of doubtful causality).

When describing the radio burst coverage with wavelength, we do not imply continuation of the same radio burst. Nevertheless, based on our observations, type III radio bursts mostly cover continuously the radio wavelength as noted in the data table. In contrast, type II radio bursts may not imply simple propagation of the driver from low corona to IP space, even though the signature is often seen both at m and km wavelengths. 

Nevertheless, the combination of given radio burst signatures at specific wavelength ranges, as well as the widely adopted origin of the probable driver, motivates us to adopt the following arguments: radio bursts observed at low corona (high frequencies, dm-wavelengths) imply acceleration at magnetic reconnection process during solar flares. Usually, type III and IV radio bursts are observed in such case. In several cases type II bursts are reported at dm$-$m but these could be also the flare-signatures of shock waves. On the other hand, DH$-$km signatures of type II bursts are well regarded as the CME-driven shocks. Over a large wavelength range, m$-$DH, both flare and CME signatures are expected to co-exist and no conclusion on the particle driver can be drawn. A partial analysis over SC24 is briefly discussed in \cite{2017simi.conf...19M}.

The following scenarios for the dominant radio burst driver are proposed based on the entire event sample:
\begin{description}
 \item[{\bf Flare acceleration}] We require the presence of type III bursts extending into the IP space, namely observed from highest (dm and dm$-$m range) to lowest frequencies (km-wavelengths), without the presence of any type II bursts (namely, we exclude shock signatures). We find only 3\% (13/431) of the entire sample that fulfill this condition. When relaxing the condition to onset of the type III radio burst also from lower wavelengths, again without any type IIs, we could add 16\% (69/431) to the above result. Thus, a dominant-to-plausible flare acceleration of radio bursts (and also SEP events) is obtained in 19\% of all cases. 
 \item[{\bf CME acceleration}] If we require CME-driven events, a focus on the presence of type II solar radio bursts is set. Since type III bursts always exist at lower frequencies, in order to minimize the flare contribution all cases with a presence of dm and dm$-$m type III bursts are omitted. Firstly, type II bursts only in DH$-$km range are considered and about 10\% (45/431) of the events comply with the requirement. When we allow the type II burst to be present also at m and/or m$-$Dm wavelengths, additionally $\sim$32\% (136/431) are found. Thus the dominant-to-probable CME contribution to solar radio bursts (and SEPs) is about twice as large ($\sim$42\%) compared to the flare-origin sample. However, some contamination from flare-produced type III in the m-to-km range cannot be entirely excluded.
 \item[{\bf Mixed contribution}] For about one third of all events, $\sim$32\% (139/431), type IIs and dm and/or dm$-$m type IIIs occur simultaneously on the dynamic radio spectra. The individual contribution of flares and CMEs cannot be clearly isolated and thus both particles accelerators could play a role.
 \item[{\bf Uncertain cases}] Events with occasional data gaps in the m-to-km wavelength ranges, constitute about 7\% (29/431) of the sample.
\end{description}

\section{Discussion and conclusion}
\label{S-Discussion}

In this work we performed a comprehensive analysis of solar radio bursts of type II, III, and IV related to proton-producing flares and CMEs during the period 1996$-$2016. Type III bursts are the most numerous radio burst type at longest wavelength regime (low frequencies, high corona/IP space) that is always present, followed by type II bursts ($\sim$60$\pm$5\%) and type IV ($\sim$15$\pm$8\%). In general, the occurrence of type II and III burst is low, if any, at dm-range and increases with the increase of radio wavelength, which is more pronounced for type III compared to the type II bursts. As argued in \cite{2013SoPh..282..579M} the low fraction of dm-bursts is due to the unfavorable conditions for the production of radio emission in the low corona. In contrast, type IV events are more numerous at dm-range (30$-$50\%), then they reach a maximum occurrence at m or/and m$-$Dm wavelengths and decrease abruptly in the IP space. Instrumental limitations, in addition to physical processes, as described in \cite{2007ApJ...671..894T}, may be the reason for the low occurrence of DH$-$km type IVs in high corona/IP space. Larger values in the DH-range, 75\% for IIs and 35\% of type IVs, are reported by \cite{2016JSWSC...6A..42P} based on slightly different yearly coverage (1997$-$2013), whereas a lower value, 83\%, is given for type IIIs, although the latter is within the uncertainty based on our analysis. In a previous study in SC23 \citep{2013SoPh..282..579M} a larger fraction by about $\sim$20\% of type DH-II bursts is reported, compared to the re-evaluated sample presented by us, whereas consistent results are obtained for type DH-IIIs and m-IVs. In the ongoing SC24 we obtain an increase to 73$\pm$10\% for DH-km type IIs, whereas the trends for the other burst types are as in SC23. No statistical validity intervals are given by earlier studies. The intermittent/patchy appearance of type IIs at DH-km wavelengths and the different period of analysis could be the reason for the reported large discrepancies, compared to the other burst types.

Generally, when considering the entire period (over almost two solar cycles), no statistical differences are obtained for the fractions of the different bursts when the proton sample is divided according to the helio-location of their solar origin. If we focus on the solar origin characteristics, Eastern-sample flares are of larger class, compared to Western flares, that could be explained as a mechanism of selection due to unfavorable (eastern) solar source location. The protons produced by weak eastern flares are not detected at Earth. CMEs show no clear tendency for faster speeds when originating from eastern location, with a larger spread in speeds than for western-directed CMEs. The speed of CMEs (regarded as a proxy for the the shock speed) related to type Western-II bursts is similar to the CME speed of the type III sample, even at the DH range (taking also into account the large uncertainties of several hundred of km$\,$s$^{-1}$), where contribution from shock-dominated acceleration is expected to prevail. Slightly larger difference in speed is noticed for the Eastern samples of type IIs and IIIs. The trends suggest that whereas western-CMEs represent a constant background source for particle (and radio burst) production, eastern-CMEs increase their influence at longer wavelengths.

Often, earlier reports \citep{{2003GeoRL..30.8013G},{2004ApJ...605..902C},{2007ApJ...658.1349C}} considered major proton events, with peak intensity larger than a specific value. For comparison, we also separated the entire sample into two parts: proton events with intensities larger or smaller than the median value for the sample. The radio burst occurrence was evaluated and it was found that, as a rule, the occurrence rate of solar radio burst types related to the strong sample is higher compared to the entire sample. This could be the source of larger fraction of type IIs reported by others. In addition, radio bursts in relation to larger proton events have statistically larger flares and faster CMEs (median values). This finding, however, can be also explained by the Big Flare Syndrome \citep{1982JGR....87.3439K} hypothesis, which speculates about the co-presence of multitude of phenomena during large eruptions without the requirement of causality between these events.

\begin{acknowledgements}
      We acknowledge the open access policy to dynamic spectral plots and event reports provided by the solar radio observatories. R.M. acknowledges support from the project `The origin of solar energetic particles: solar flares vs. coronal mass ejections', co-funded by the National Science Fund of Bulgaria contract No. DNTS/Russia 01/6 (23-Jun-2017) and the Russian Foundation for Basic Research Project No. 17-52-18050. V.K. acknowledges support by an appointment to the NASA postdoctoral program at the NASA Goddard Space Flight Center administered by Universities Space Research Association under contract with NASA and the Czech Science Foundation grant 17-06818Y.
\end{acknowledgements}

\bibliography{mybibfile}


\end{document}